\documentclass[twocolumn,english]{revtex4-1}
\usepackage[T1]{fontenc}
\usepackage[latin9]{inputenc}
\usepackage{booktabs}
\usepackage{amsmath}
\usepackage{amssymb}
\usepackage{graphicx}

\makeatletter

\providecommand{\tabularnewline}{\\}

\usepackage{babel}
\usepackage{color}

\makeatother

\begin{document}
\title{Exchange interactions from a nonorthogonal basis set:
from bulk ferromagnets to the magnetism in low-dimensional graphene systems
}
\author{L\'aszl\'o Oroszl\'any$^{1,2}$}
\author{Jaime Ferrer$^{3,4}$}
\author{Andr\'as De\'ak$^{5,6}$}
\author{L\'aszl\'o Udvardi$^{5,6}$}
\author{L\'aszl\'o Szunyogh$^{5,6}$}
\affiliation{$^{1}$Department of Physics of Complex Systems, E\"otv\"os Lor\'and University,
P\'azm\'any P\'eter s\'et\'any 1/A, H-1117 Budapest, Hungary}
\affiliation{$^{2}$MTA-BME Lend\"ulet Topology and Correlation Research Group, Budapest
University of Technology and Economics, Budafoki \'ut 8., H-1111 Budapest,
Hungary}
\affiliation{$^{3}$Departamento de Fisica, Universidad de Oviedo, 33007 Oviedo,
Spain}
\affiliation{$^{4}$Nanomaterials and Nanotechnology Research Center (CINN), CSIC
- Universidad de Oviedo}
\affiliation{$^{5}$Department of Theoretical Physics, Budapest University of Technology
and Economics, Budafoki \'ut 8., H-1111 Budapest, Hungary }
\affiliation{$^{6}$MTA-BME Condensed Matter Research Group, Budapest University
of Technology and Economics, Budafoki \'ut 8., H-1111 Budapest, Hungary }
\begin{abstract}
We present a computational method to determine the exchange constants
in isotropic spin models. The method uses the Hamiltonian and overlap
matrices computed from density functional schemes that are based on
nonorthogonal basis sets. We demonstrate that the new method as implemeted in the SIESTA code reproduces
the Heisenberg interactions of simple metallic bulk ferromagnets
as obtained from former well--established computational approaches.
Then we address $sp$ magnetism in graphene nanostructures.
For fluorinated graphene we obtain exchange interactions in fairly good
agreement with previous calculations using maximally localized Wannier functions
and we confirm the theoretical prediction of a 120$^\circ$ N\'eel state.
Associated with the magnetic edge-states of a zigzag graphene nanoribbon we find
rapidly decaying exchange interactions, however, with an unconventional distance
dependence of $\exp(-\sqrt{r/\delta})$.
We show that the stiffness constant derived from the exchange interactions is
consistent with previous estimate based on total energy differences of twisted
spin configurations.
We highlight that our method is an efficient tool
for the analysis of novel hybrid nano-structures where metallic and
organic components are integrated to form exotic magnetic
patterns.
\end{abstract}
\maketitle

\section{Introduction}

Heisenberg-like spin Hamiltonians form a solid basis for describing
the ground state and thermal behavior of a wide range of magnetic systems, either
characterized by itinerant electrons or by local moments. The exchange
constants entering the spin model can be derived from first principles.
One of the most frequently used approaches is based on the magnetic force theorem
originally introduced by Liechtenstein and co-workers \cite{Lichtenstein_original},
referred to as the LKAG formalism as what follows.
Related methods have been developed since then to tackle correlated
systems \cite{Katsnelson_Jij_TBcorr,KatsnelsonEPJB2002}, relativistic
effects \cite{Udvardi_relativistic_Jij,ebertPRB09} or both of them
\cite{Katsnelson_Jij_TBcorr,SecchiAnnPhys2015}.

The use of one-electron Green's functions is an integral part of the
formalism in Ref.~\cite{Lichtenstein_original}. Therefore, the Korringa--Kohn--Rostoker
Green's function (KKR-GF) \cite{KKRBOOK} and the tight-binding linear
muffin-tin orbital (TB-LMTO) methods \cite{TBLMTO-Andersen,TBLMTO-Turek}
have been particularly successful for calculating magnetic exchange
interactions for bulk materials, surfaces, interfaces, films, superlattices,
and even for finite metallic clusters \cite{Ebert2011,Pajda2001,Pajda2000}.
Furthermore, the calculation of tensorial exchange interactions, including
two-ion magnetic anisotropy parameters and Dzyaloshinskyi-Moriya interactions,
has become available by extending the LKAG formula to relativistic
systems \cite{Udvardi_relativistic_Jij,ebertPRB09}. This extension
has opened the door to the analysis, design and tuning of complex
magnetic states like domain walls \cite{Vida-domainwalls}, spin spirals
\cite{Rozsa-FeWTa,Simon-cappedCoPt} and magnetic skyrmions \cite{Polesya-FeTaS-skyrmion,Rozsa-skyrmion1,Rozsa-skyrmion2,Hsu-HFePt-skyrmions}
in ultrathin films. This extension also enables us to study the recently
discovered van der Waals ferromagnets \cite{Gong17,Huang17,Bonilla18},
whose magnetic state is stabilized by the anisotropy barriers that
overcome the thermal spin fluctuations standing behind the Mermin--Wagner
theorem \cite{Mermin66}.

The KKR-GF or TB-LMTO methods are not fully adapted to describe open
systems such as atoms or molecules deposited on surfaces or suspended
in nano-scale junctions because they commonly make use of the atomic
sphere approximation (ASA). Such systems are accurately treated
by other methods like the program packages VASP \cite{vasp}, Quantum
Espresso \cite{quantespr}, SIESTA\cite{siesta} or ADF\cite{ADF2017}.
The first two expand the eigenstates onto a plane-wave basis \cite{vasp,quantespr},
so a transformation to maximally localized Wannier functions \cite{Maximally_Localized_Wannier_RMP}
is needed to determine an orthogonal tight-binding basis set requested by the LKAG
formalism \cite{Katsnelson_Jij_Wannier}.
In contrast, SIESTA and ADF conveniently use a basis set of wave functions
that are localized on each individual atom, possess orbital quantum
numbers and are nonorthogonal, hence producing self-consistent Hamiltonians
and overlap matrices that are already written in the tight-binding
language. In order to calculate exchange constants in the spirit of the magnetic
force theorem 
orthogonal basis sets, a generalization of the LKAG formula to nonorthogonal
bases is called for. This is the main goal and accomplishment of the
present article. We therefore develop a formalism which enables us
to use some popular density-functional codes to evaluate exchange
interactions of isotropic Heisenberg models in a broad range of magnetic
systems. We implement the new scheme in the SIESTA code and present
results for conventional bulk metallic magnets as well as for graphene
ribbons and fluorinated graphene sheets displaying $sp$ magnetism
that compare well with those available in the literature. We outline
the main features of our approach in Section II, then we present and discuss
our results in Section III. A short summary closes the article. All
the algebraic details of the approach can be found in the appendices.

\section{Method}

The classical Heisenberg model describes a lattice of localized classical
spins characterized by unit vectors $\vec{e}_{i}$, where $i$ denotes lattice
(or atomic) sites. Within a nonrelativistic theory, the spin-spin
interactions are encapsulated in terms of isotropic exchange parameters
$J_{ij}$ entering the spin Hamiltonian,
\begin{equation}
\mathcal{H}=-\frac{1}{2}\sum_{i\ne j}J_{ij}\,\vec{e}_{i}\vec{e}_{j}\,.\label{eq:heisemberg_model}
\end{equation}

The magnetic force theorem enables us to extract the exchange parameters
from the effective single-particle Hamiltonian $\hat{H}$ that results
from \emph{ab initio} calculations \cite{Lichtenstein_original,Udvardi_relativistic_Jij}.
We use a collinear-spin reference frame and write $\hat{H}$ in terms
of a basis set of localized orbitals centered at lattice sites. Consequently,
the tight-binding Hamiltonian matrix $\boldsymbol{H}$ and the overlap matrix
$\boldsymbol{S}$ are diagonal in the spin indices.
Collecting all the basis functions assigned to a given site $i$, the
corresponding spin-dependent and site-indexed blocks of the Hamiltonian are
denoted by $\underline{H}_{ij}^{\sigma}$ and, similarly, $\underline{S}_{ij}$
for the overlap matrix which is independent on the spin index since in practice
for both spin channels the same basis functions are considered. We then find
that the exchange parameters can be derived from the expression
\begin{equation}
J_{ij}=\frac{2}{\pi}\int_{-\infty}^{\varepsilon_\text{F}}\text{d}\varepsilon\,\text{ImTr}_{L}[\underline{H}_{ii}^{s}\underline{\tilde{G}}_{ij}^{\uparrow}(\varepsilon)\underline{H}_{jj}^{s}\underline{\tilde{G}}_{ji}^{\downarrow}(\varepsilon)]\,,\label{eq:nojij}
\end{equation}
where $\varepsilon_\text{F}$ is the Fermi energy, Tr$_{L}$ denotes the
trace of matrices in orbital space,
\begin{eqnarray}
\underline{H}_{ii}^{s} & = & \frac{\underline{H}_{ii}^{\uparrow}-\underline{H}_{ii}^{\downarrow}}{2},
\end{eqnarray}
and
\begin{eqnarray}
\underline{\tilde{G}}_{ij}^{\sigma}(z) & = & \left[\left(z\boldsymbol{S}-\boldsymbol{H}\right)^{-1}\right]_{ij}^{\sigma}
\end{eqnarray}
is the appropriate site off-diagonal block of the matrix of expansion
coefficients of the resolvent operator $\hat{G}(z)=\left(z\hat{I}-\hat{H}\right)^{-1}$
with $z=\varepsilon+\text{i}\delta$.
A detailed derivation of Eq.~\eqref{eq:nojij}
is given in the appendices. These expressions are a generalization
of the seminal work of Liechtenstein {\em et al.}~\cite{Lichtenstein_original}
to the case of a nonorthogonal tight-binding basis set.

We implemented the above equations by using the self-consistent Hamiltonian
and overlap matrices provided by the SIESTA code \cite{SIESTA_paper}.
This can be achieved with the assistance of the \texttt{sisl}
tool \cite{sisl}. We devote the next section to validate our approach
by giving three examples \cite{nojijrepo} by comparing the results of our proposed
methodology with previous calculations. Moreover, we give a detailed description of the $sp$ magnetism in low-dimensional graphene systems in terms of exchange interactions and analyze their asymptotic behavior.

\section{Results}

\subsection*{Bulk ferromagnets}

In this section we present the exchange interactions of selected bulk
ferromagnets using our proposed approach and compare our results with
former ones obtained from the screened KKR (SKKR) method \cite{KKRBOOK}
in the framework of the atomic sphere approximation. We considered
ferromagnetic bcc Fe, hcp Co as well as fcc Ni. For both the SIESTA and SKKR
calculations we used the same approximations for the exchange-correlation
density functional and the same geometrical parameters. In case of
Fe the generalized gradient approximation (GGA) as parameterized
by the PBE scheme \cite{PBE}, while for Co and Ni the local spin density
approximation (LSDA) \cite{PZ-LSDA} were employed. The lattice constants
$2.87\,\text{\AA}$, $2.50\,\text{\AA}$ and $2.49\,\text{\AA}$ were chosen for
bcc Fe, hcp Co and fcc Ni, respectively. In addition, for Co we considered the
ratio of $c/a=1.633$ of an ideal hcp structure.
scheme. However, we found that choosing a $k$-space cutoff of $50\,\text{\AA}^{-1}$
and a real space mesh cutoff of at least $500$~Ryd ensured reliable accuracy
for the SIESTA results. We noticed, however, that the choice of the
pseudo-potential parameters had a considerable impact on the results
for the ground state obtained from SIESTA and, subsequently,
also on the calculated exchange parameters. In our calculations we
used the pseudo-potential generation scheme described in Ref.~\cite{Jaime_Pseudos}.

The calculated spin magnetic moments of the three bulk ferromagnets are
summarized in Table \ref{tab:moments_for_elementary magnets} for the two
self-consistent schemes. The data in this table show an almost perfect agreement
between the spin moments obtained from the two \emph{ab initio} methods for bcc
Fe, and relative differences of about 3 \% and 8
\% for hcp Co and for fcc Ni, respectively.

\begin{table}[ht]
\caption{\label{tab:moments_for_elementary magnets}Spin magnetic moments in
units of $\mu_\text{B}$ for the bulk ferromagnets under consideration, as
calculated using the SKKR method and the SIESTA code.}
\begin{centering}
\begin{tabular*}{0.45\columnwidth}{@{\extracolsep{\fill}}@{\extracolsep{\fill}}ccc}
\toprule
 & SKKR  & SIESTA\tabularnewline
\midrule
\midrule
bcc Fe  & 2.365  & 2.356\tabularnewline
\midrule
hcp Co  & 1.542  & 1.580 \tabularnewline
\midrule
fcc Ni  & 0.675  & 0.626\tabularnewline
\bottomrule
\end{tabular*}
\par\end{centering}
\end{table}

Next we calculated the isotropic exchange parameters for the three bulk
ferromagnets by using the relativistic torque method within the SKKR
\cite{Udvardi_relativistic_Jij} and via the formula in Eq.~\eqref{eq:nojij} with the
 tight-binding Hamiltonian and overlap matrices obtained
from SIESTA. It should be mentioned that in the latter case we needed a
$100$ $k$-point mesh in each direction of the full Brillouin zone to ensure
adequate convergence  for the real space Green's function expansion
coefficient matrices $\underline{\tilde{G}}_{ij}^{\sigma}$.

\begin{figure}[ht]
\includegraphics[width=1\columnwidth]{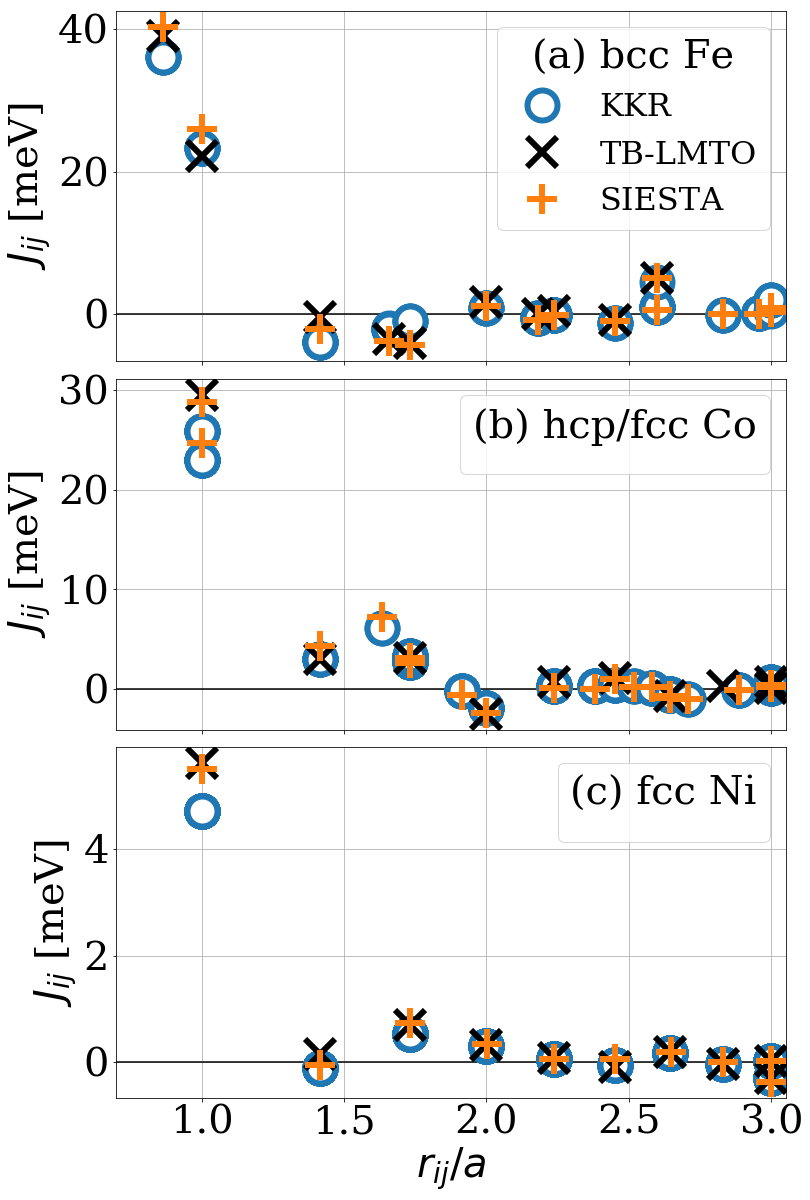}

\caption{\label{fig:elementary_magnets_jij} (color online) Isotropic exchange
interactions $J_{ij}$ as a function of inter-site distances $r_{ij}$ measured
in units of the lattice constants $a$, calculated using three different
computational approaches: (a) bcc Fe; (b) hcp Co from SIESTA and SKKR, fcc Co
from TB-LMTO; and (c) fcc Ni. Blue circles, black $\times$ and  orange $+$ symbols
stand for the SKKR, TB-LMTO \cite{Pajda2001} and  SIESTA calculations,
respectively.}
\end{figure}

The exchange parameters $J_{ij}$ obtained from the two methods are plotted in
Fig.~\ref{fig:elementary_magnets_jij}. For a more extensive comparison we also
included the corresponding values reported in Ref.~\cite{Pajda2001} in terms of
the TB-LMTO approach. Note that the spin model considered in Ref.~\cite{Pajda2001}
misses the factor of $1/2$ in Eq.~\eqref{eq:heisemberg_model}, therefore, the
exchange interactions presented there should be multiplied by a factor of $2$ in
order to compare with those calculated from Eq.~\eqref{eq:nojij}. Apparently,
the three methods provide isotropic exchange interactions in remarkably good
agreement with each other for all three bulk ferromagnets. Considering mainly
the large ferromagnetic nearest neighbor interactions, but also in case of some
farther couplings, the SIESTA and TB-LMTO values compare more precisely than
those and the SKKR values, which is not surprising as the former two methods
rely on the tight-binding scheme. As can be seen in Fig.~\ref{fig:elementary_magnets_jij}(b),
the exchange interactions derived from the SIESTA and SKKR calculations also
compare remarkably well with those obtained from the TB-LMTO method for fcc Co.

The Curie temperature $T_{\rm C}$ of ferromagnetic materials is one of the
measurable quantities closely related to the exchange interactions.  While the
transition temperature can accurately be obtained from Monte Carlo or
spin-dynamics simulations, here we present theoretical estimates based on the
mean field approach which is extracted from the spin model parameters $J_{ij}$
as
\begin{equation}
T^{\rm MFA}_{\rm C}=\frac{1}{3 k_{\rm B}}\sum_{j}J_{0j},
\label{eq:tcmf}
\end{equation}
with the Boltzmann constant $k_{\rm B}$. We calculated $T^{\rm MFA}_{\rm C}$
summing up the exchange parameters up to a distance of $r_{0j}=10$ \AA  \ for
hcp Co and fcc Ni, while $r_{0j}=25$ \AA \ for bcc Fe,  reducing the numerical
error of the results below 20 K. The data obtained within the SIESTA and SKKR
methods shown in Table \ref{tab:Tc} are in fairly good agreement with each other
and with those reported in Ref.~\cite{Pajda2001}, also presented in Table \ref{tab:Tc}.
The somewhat large deviation of  $T^{\rm MFA}_{\rm C}$  of Co within the TB-LMTO
method from the very similar values obtained using the SIESTA and SKKR codes
can mainly be attributed to the different crystal structures used in these
calculations. The mean-field approximation is known to overestimate the exact
transition temperatures, which might explain the higher values of $T^{\rm MFA}_{\rm C}$
as compared with the experimental $T_{\rm C}$ in case of Fe and Co. The
considerably lower mean-field estimates for the Curie temperature with respect
to the experimental value in case of Ni is most possibly the consequence of the
highly itinerant nature of the magnetism of bulk Ni \cite{Staunton_1992,Staunton_2014}.


\begin{table}[ht]
\caption{\label{tab:Tc}Mean-field Curie temperatures $T^{\rm MFA}_{\rm C}$ for elementary ferromagnets calculated using Eq.~\eqref{eq:tcmf} with exchange parameters obtained from different \emph{ab initio} methods. Note that the TB-LMTO results for Co correspond to an fcc structure. For comparison, experimental Curie temperatures are also presented in the last column.}
\begin{tabular*}{1\columnwidth}{@{\extracolsep{\fill}}@{\extracolsep{\fill}}ccccc}
\toprule
 & SKKR  & TB-LMTO\cite{Pajda2001}  & SIESTA  & Experiment\cite{Mook-Pajda-36,Pauthenet-Pajda-37-1,Pauthenet-Pajda-37-2,Shirane-Pajda-38}\tabularnewline
\midrule
\midrule
bcc Fe  & 1478  & 1414  & 1330  & 1044-1045\tabularnewline
\midrule
hcp Co  & 1504  & 1645  & 1490  & 1388-1398\tabularnewline
\midrule
fcc Ni  & 348  & 397  & 389  & 624-631\tabularnewline
\bottomrule
\end{tabular*}
\end{table}

\subsection*{Fluorinated graphene}

\begin{figure}
\includegraphics[width=1\columnwidth]{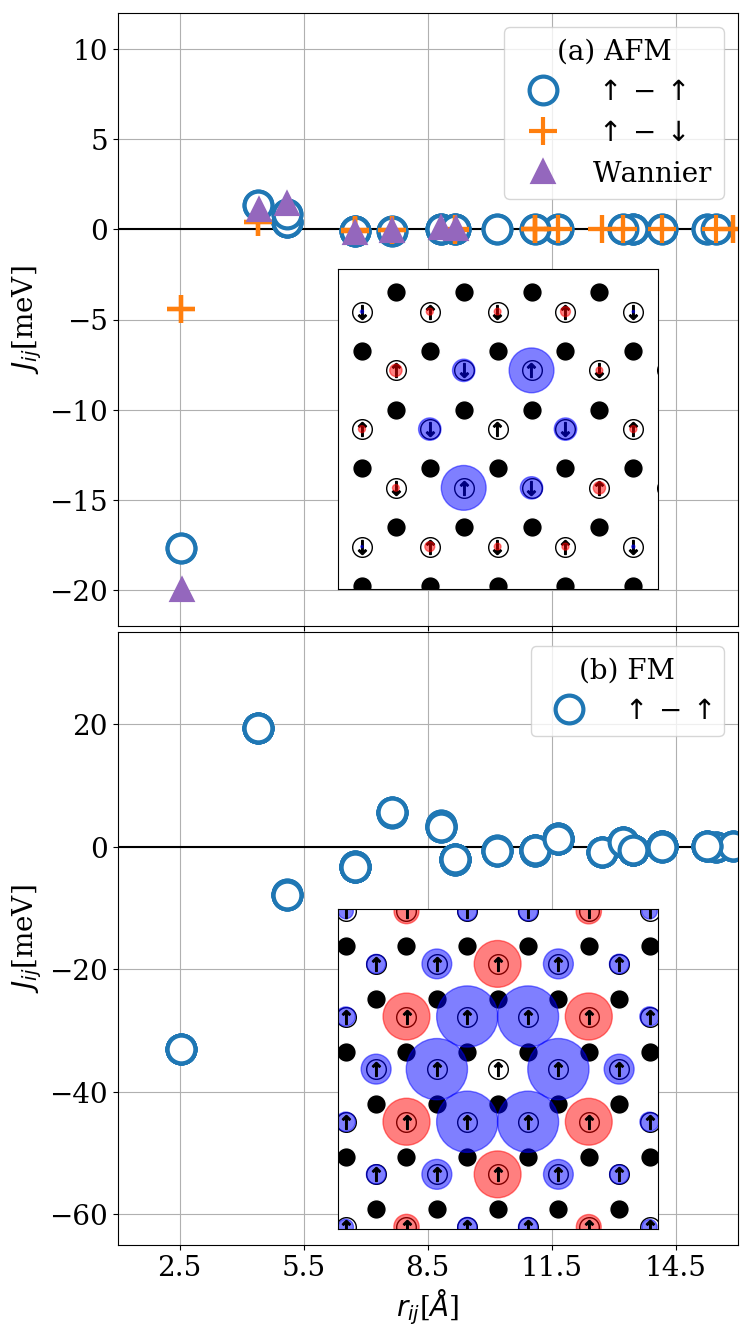}
\caption{\label{fig:C2F_Wannier_vs_SIESTA} (color online) (a) Isotropic exchange
interactions $J_{ij}$  for a fluorinated graphene sample calculated using a Wannier basis \cite{Katsnelson_Jij_Wannier}
(triangles) and extracted from a nonorthogonal basis using the SIESTA
code. For both calculations a row-wise antiferromagnetic spin configuration was used as reference. Circles and crosses
represent couplings between spins of parallel and anti-parallel orientations,
respectively. The inset shows the atomic arrangement of the carbon
atoms in this system. Here, black circles label carbon atoms bound
to a fluorine atom. Arrows in the white circles represent the orientation
of local moments, displaying a row-wise antiferromagnetic state.
Color-coded disks denote the relative magnitude and sign (red positive,
blue negative) of the exchange couplings between the central site
and other magnetic carbon sites. (b) Exchange interactions $J_{ij}$
calculated from a nonorthogonal basis set using the SIESTA code and
a ferromagnetic state as reference. The inset demonstrates that in this case the
exchange parameters respect the symmetry of the underlying
lattice. }
\end{figure}

We turn now to $sp$ magnetism in the context of graphene.
First we present results for the exchange interactions in single-side
fluorinated graphene $\text{C}_{2}\text{F}$ and compare them to earlier
calculations by Rudenko {\em et al.} \cite{Katsnelson_Jij_Wannier},
who used a maximally localized Wannier function basis  \cite{Maximally_Localized_Wannier_RMP} which
was mapped from a plane wave basis \cite{Quantum_Espresso}. Wannier
orbitals form an orthonormal basis representation, thus Eq.~\eqref{eq:Lichtenstein_TB}
can simply be evaluated using the corresponding matrices with respect
to this representation. However, Wannier orbitals are not necessarily
localized to a single atom. Therefore, local degrees of freedom like
the atomic spin can not be unambiguously described in terms of a Wannier
basis. In our approach, every nonorthogonal orbital is explicitly
localized to a given atom in the system. The nonorthogonality of these orbitals
can be handled by using appropriate local projection operators, as
discussed in Appendix B. Hence, these orbitals describe properly atomic
degrees of freedom.

In Ref.~\cite{Katsnelson_Jij_Wannier} it was found that a row-wise
antiferromagnetic (AFM) spin alignment is preferred with respect to
the ferromagnetic (FM) state. In our self-consistent calculations
performed with SIESTA we also considered a row-wise AFM configuration.
For better comparison, we used the exchange-correlation functional and geometry
parameters of Ref.~\cite{Katsnelson_Jij_Wannier}. Note that the F atoms are placed above the C atoms in
only one of the two sublattices of graphene (say, in sublattice A),
forming thus a triangular lattice. We found that the carbon
atoms at sublattice B have a total magnetic moment of $0.736\,\mu_\text{B}$,
with a contribution of $0.65\,\mu_\text{B}$ coming from their $p_{z}$
orbitals and that the carbon atoms at sublattice A have negligible magnetic moments.
This is in good agreement with the results of Ref.~\cite{Katsnelson_Jij_Wannier},
where considerable spin polarization was found only for the $p_{z}$
type Wannier orbitals associated with the carbon sites at the B sublattice,
with a magnitude of $0.59\,\mu_\text{B}$.

\begin{figure}
\includegraphics[width=1\columnwidth]{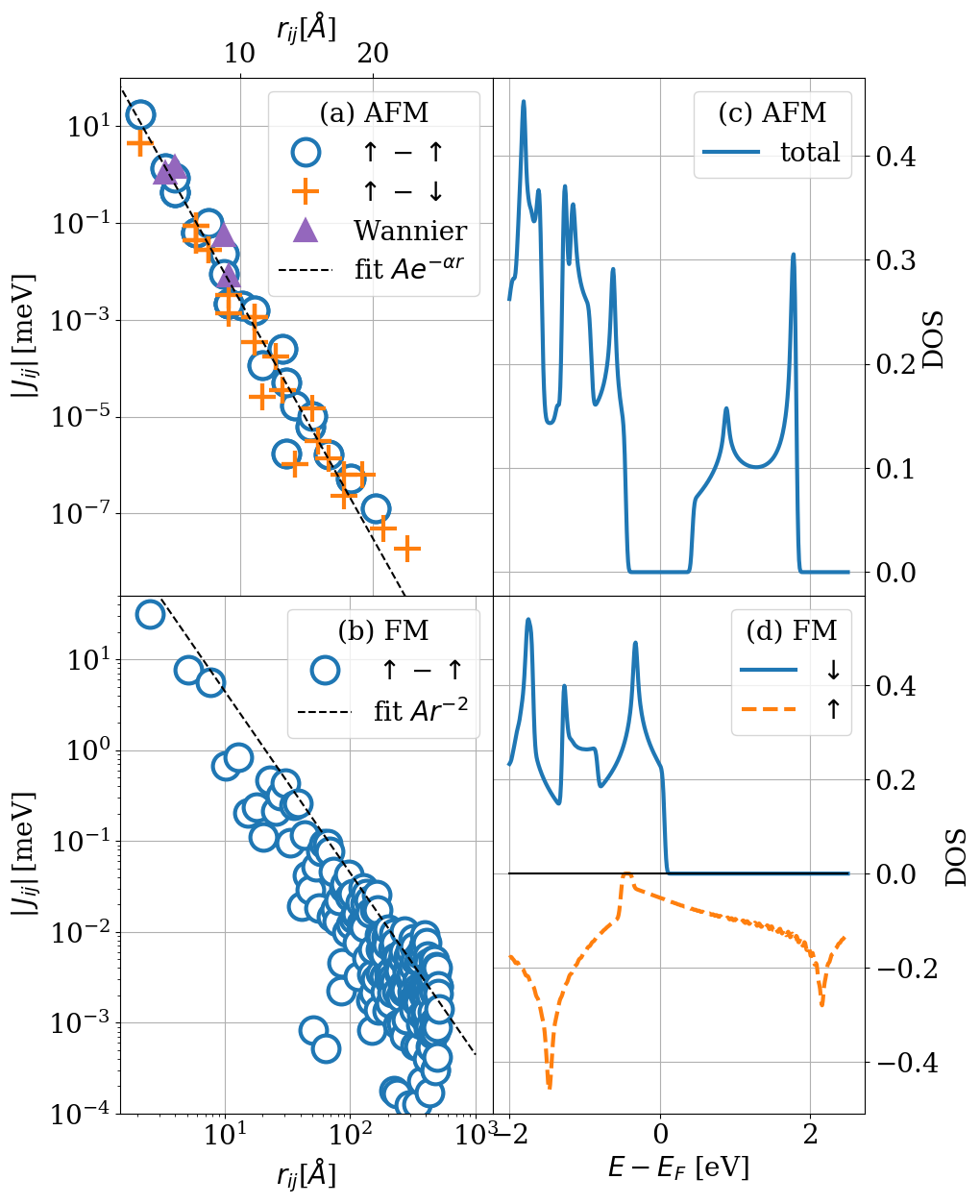}
\caption{\label{fig:C2F_log_DOS} (color online) (a) and (b) Magnitudes of the
exchange interactions between magnetic carbon atoms as a function of the
distance in fluorinated graphene in the row-wise AFM and in the FM states,
respectively. Blue circles denote interactions between atoms with the
same magnetic orientation, orange crosses denote interactions between moments
with opposite orientation, while purple triangles stand for the interactions
reported in Ref.~\cite{Katsnelson_Jij_Wannier}. Note that panel (a) is plotted
on a log-linear scale, while panel (b) on a log-log scale. The total density of
states for the AFM state and the spin-resolved densities of states for the FM
state in the vicinity of the Fermi level are plotted in panel (c) and (d),
respectively.}
\end{figure}

Choosing the row-wise AFM configuration as a reference, we calculated
the exchange parameters of $\text{C}_{2}\text{F}$ by using Eq.~\eqref{eq:nojij}
with $\underline{\tilde{G}}_{ij}^{\sigma}$-s evaluated on a $k$-mesh
of 200 points in each direction of the two-dimensional Brillouin zone.
The resulting exchange interactions are shown in Fig.~\ref{fig:C2F_Wannier_vs_SIESTA}(a). Note that we label the interactions between the moments of the
same and opposite orientations with different symbols.
Since a row-wise AFM spin configuration does not respect the point
symmetry of the triangular lattice, these two sets of interactions
significantly differ from each other: the first nearest neighbor interactions
between moments with the same orientation are much stronger antiferromagnetic
than those between opposite moments. Notably, the exchange interactions reported
in Ref.~\cite{Katsnelson_Jij_Wannier}, labeled by triangles in Fig.~\ref{fig:C2F_Wannier_vs_SIESTA}(a),
show a good agreement with those calculated by SIESTA between moments of the same orientation at
positions along a row of the row-wise AFM pattern. Moreover, farther
couplings decrease exponentially as shown in Fig.~\ref{fig:C2F_log_DOS}(a). This can be understood as a consequence of the quasiparticle
gap that is manifested in the total density of states shown in Fig.~\ref{fig:C2F_log_DOS}(c).

As mentioned above, the chosen reference state and, consequently, the
calculated exchange parameters do not respect the $C_{3v}$ symmetry
of the underlying lattice. For this reason, we also chose
the ferromagnetic state shown as an alternative reference state, and we
re-calculated the exchange parameters. The results are shown in Fig.~\ref{fig:C2F_Wannier_vs_SIESTA}(b).
We found that the exchange parameters are now consistent with the $C_{3v}$
symmetry of the lattice and that they are characterized by large
antiferromagnetic first nearest neighbor and ferromagnetic next-nearest neighbor
couplings. The decay of the exchange interactions shows the expected power law
$\propto r^{-2}$ \cite{FischerKleinRKKY} as depicted in Fig.~\ref{fig:C2F_log_DOS}(b)
and is corroborated by the absence of a gap in the density of states that we
show in Fig.~\ref{fig:C2F_log_DOS}(d).

\begin{figure}
\includegraphics[width=1\columnwidth]{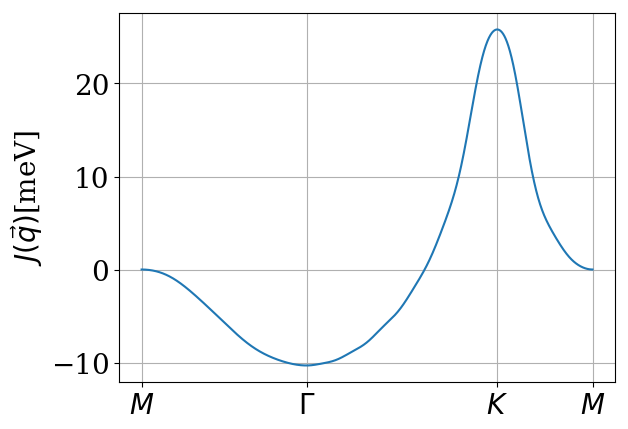}
\caption{\label{fig:C2F_Jq} The Fourier transform $J(\vec{q})$ of the exchange
constants of the fluorinated graphene sample extracted from the ferromagnetic configuration plotted along high symmetry directions of the Brillouin
zone.}
\end{figure}

It should be mentioned that both the first nearest neighbor antiferromagnetic
and second nearest neighbor ferromagnetic couplings are consistent
with the 120$^{\circ}$ N\'eel ground state suggested in Ref.~\cite{Katsnelson_Jij_Wannier}.
We evaluated the Fourier transform $J(\vec{q})$ of the exchange interactions
obtained from the ferromagnetic reference state and plotted it in Fig.~\ref{fig:C2F_Jq}
along the high symmetry directions of the hexagonal Brillouin zone. Indeed we
obtain a clear maximum 
at the $K$ point, which indicates the 120$^{\circ}$ N\'eel state as ground state.

\subsection*{Graphene ribbon}

\begin{figure*}[ht]
\includegraphics[width=0.8\textwidth]{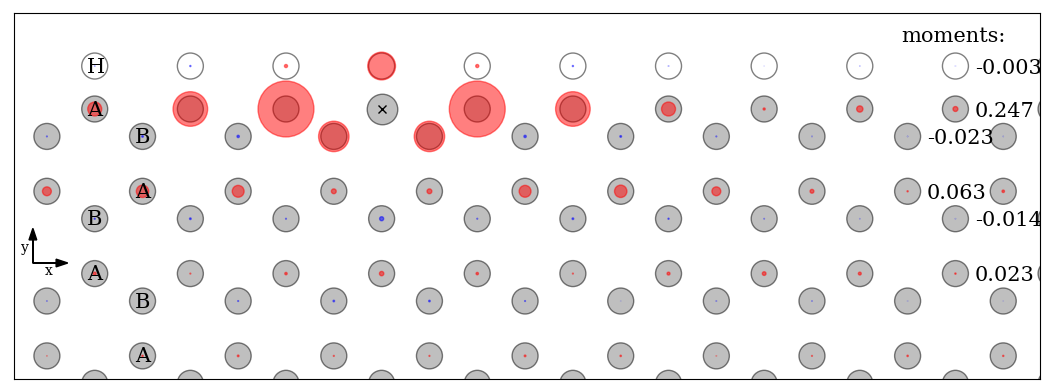}
\caption{\label{fig:graphene_ribbon} (color online) Atomic arrangement at
the edge of a hydrogen-passivated zigzag graphene ribbon. White circles
denote the passivating hydrogen atoms, while gray circles denote carbon
atoms on sub-lattices A and B. The sub-lattice of a given row is indicated
at the left side of the figure, while the value of the local magnetic
moment in the row is shown at the right side in units of $\mu_\text{B}$.
Color-coded disks denote the relative magnitude and sign (red positive,
blue negative) of the $J_{ij}$ exchange interactions between the
carbon atom marked by $\times$ and the corresponding sites.}
\end{figure*}

In this section we analyse
the one-dimensional itinerant ferromagnetic state that arises
at the edge of zigzag graphene ribbons passivated by hydrogen atoms \cite{Fujita96,Son06,Meunier17,Louie11,Magda2014,Hagymasi2016}.
This rather elusive magnetic state has provoked enormous expectations
in the past fifteen years, because of their plausible potential for
spintronics applications \cite{Slota2018,Pesin12}.

We considered a hydrogen-passivated $28$-carbon-atom wide graphene
ribbon that extends along the $x$ direction, as depicted in Fig.~\ref{fig:graphene_ribbon}.
In the self-consistent calculations we used the LSDA \cite{PZ-LSDA},
we set a mesh cutoff of $200$ Ry for the real-space integrals and
we selected $100$ $k$-points in the one-dimensional (1D) Brillouin-zone. The
magnetic configuration was set ferromagnetic along the $x$ direction, that means
the 1D unit cell consisted of 28 carbon atoms and 2 H atoms.
 The magnetic moments yielded by this choice are indicated in Fig.\ \ref{fig:graphene_ribbon}.
 We find that the magnetic moments of the A and B carbon atoms are opposite in
 sign, implying that the ribbon's ground state displays  an antiferromagnetic
 alignment with respect to the two edges.
 A total magnetic moment of $0.3\,\mu_\text{B}$ per 1D unit cell is associated
 with each edge, that we obtain by adding the individual magnetic moments of the
 ribbon atoms from one edge to the center of the ribbon. The two sublattices A
and B contribute $0.36\,\mu_\text{B}$ and $-0.06\,\mu_\text{B}$ to this
magnetic moment, respectively.
These findings are in good agreement with previous calculations \cite{Katsnelson_graphene_ribbon,Jaime_ribbon}.

\begin{figure}[ht]
\includegraphics[width=1\columnwidth]{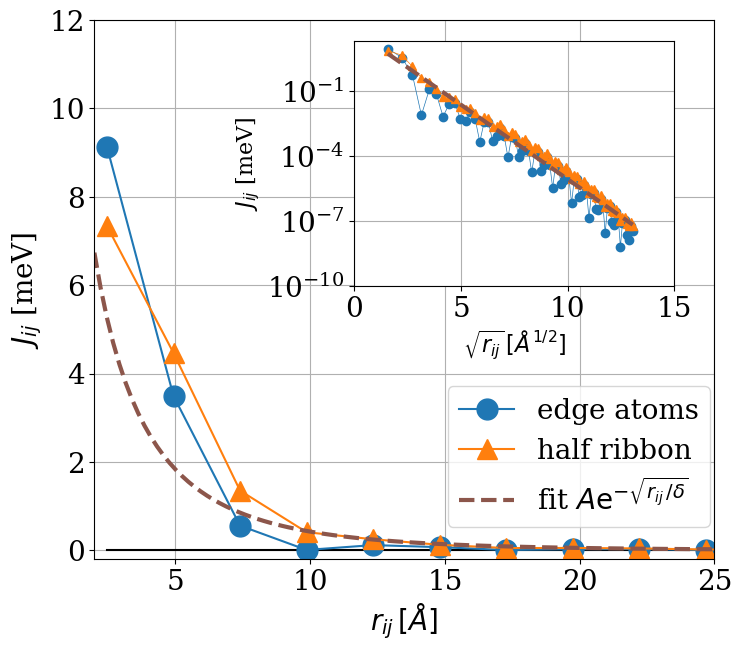}
\caption{\label{fig:graphene_ribbon_Jij} (color online) Isotropic exchange
interactions $J_{ij}$ calculated for a graphene ribbon as a function
of the $x$ coordinate measured along the edge. Blue circles correspond
to the interactions between carbon atoms of type A located in the row 
at the edge, while orange triangles denote exchange interactions between
meta magnetic moments comprising all the atoms from one half of the cross section
of the ribbon. The dashed line is a fitting of the tail of the latter curve to the function
$A\,e^{-\sqrt{r_{ij}/\delta}}$ . A log-linear
graph of both $J_{ij}$ curves as a function of the square root of the distance
is depicted in the inset.}
\end{figure}

We calculated the exchange parameters $J_{ij}$ in the ribbon using the LKAG
formula Eq.~\eqref{eq:nojij} with the ground-state Hamiltonian delivered by
SIESTA. We  found that the leading interactions occur between those edge carbon
atoms on sublattice A that have the largest magnetic moments.
Fig.~\ref{fig:graphene_ribbon} demonstrates that these interactions are
ferromagnetic and fairly short-ranged. The decay of the interactions is
non-monotonic: a small oscillatory behavior can be observed, but the
interactions remain ferromagnetic for all distances.
Our calculations also indicate non-negligible antiferromagnetic couplings
between the A atoms at the edge and the first nearest neighbor atoms at the B
sublattice.

In order to make a conceptional connection to the calculation presented
in \cite{Katsnelson_graphene_ribbon}, we introduce meta magnetic
moments as large as $0.3\,\mu_\text{B}$ associated with one half of the graphene
ribbon. To calculate the interactions between these meta moments we evaluate
Eq.~\eqref{eq:nojij} considering all atomic orbitals in one half of the 1D unit
cell of the ribbon containing 14 carbon atoms and one hydrogen atom. In
Fig.~\ref{fig:graphene_ribbon_Jij} we also plot these exchange interactions as a
function of the distance along the $x$ direction by orange triangles.
The nearest-neighbor interaction between the meta moments is somewhat reduced,
while the farther interactions are enhanced compared to the corresponding
interactions between the edge atoms, as can be seen in Fig.~\ref{fig:graphene_ribbon_Jij}.
Consequently, the interactions between the meta moments show a fairly monotonic
decay, with similar characteristics as in case of the edge atoms for distances
above $15$~\AA . We found that in this region the interactions fit well to a
function $\propto\text{e}^{-\sqrt{r/\delta}}$, also shown in Fig.~\ref{fig:graphene_ribbon_Jij}.
In order to better visualize this unconventional decay we plotted the exchange
parameters on a logarithmic scale against the square root of the distance in the
inset of Fig.~\ref{fig:graphene_ribbon_Jij}. Both sets of interactions display a
nearly linear dependence on this graph, though the aforementioned oscillations
 for the edge atoms clearly show up.

\begin{figure}
\includegraphics[width=1\columnwidth]{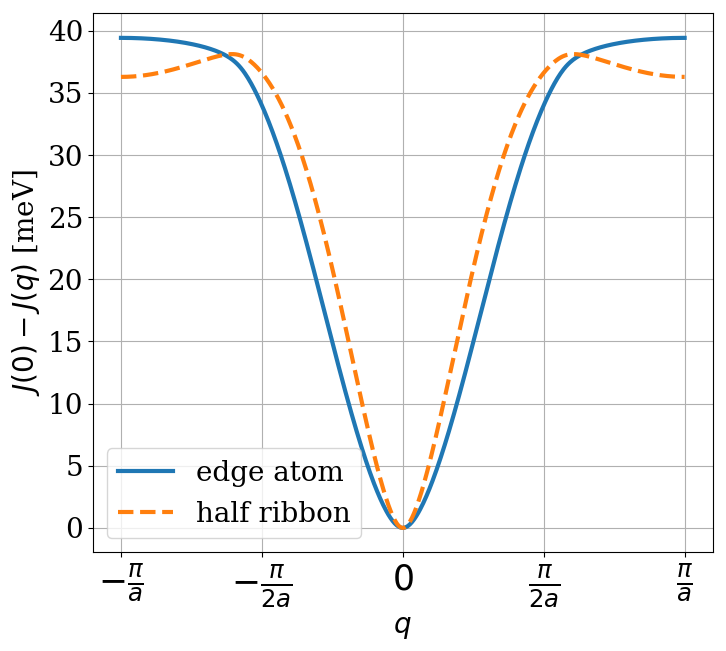}
\caption{\label{fig:graphene_ribbon-magnon} (color online) Spin-wave spectrum
(without the prefactor $\frac{2\mu_\text{B}}{M}$) for the magnetic moments at
the edge of the graphene ribbon only (solid blue line) and for the meta magnetic
moments corresponding to half of the ribbon (orange dashed line).}
\end{figure}

The experimentally accessible magnon spectrum $E(q)$ of a ferromagnetic system
is related to the Fourier transform of the exchange constants  \cite{Pajda2001},
\begin{equation}
E(q)=\frac{2\mu_\text{B}}{M}\left(J(0)-J(q)\right) ,
\end{equation}
where $M$ is the magnitude of the magnetic moment per periodic unit.
The calculated curves for $J(0)-J(q)$ for the two considered spin models are
plotted in Fig.~\ref{fig:graphene_ribbon-magnon}, clearly proving that the
ferromagnetic state ($q=0$) is the ground state of the half ribbon system.

The low-energy magnon spectrum
\begin{equation}
E(q)\approx Dq^{2}
\end{equation}
of the ribbon was estimated in Ref.~\cite{Katsnelson_graphene_ribbon}
from a set of constrained self-consistent calculations of twisted periodic spin
configurations resulting in a stiffness constant of $D=2100\,\text{meV\AA}^{2}$.
Based on this value of $D$, an effective first nearest neighbor Heisenberg model
was devised with the exchange constant $J_{01}=105\,\text{meV}$.

The stiffness constant $D$ can be calculated from the exchange parameters as
\begin{equation}
D=\frac{\mu_\text{B}}{M}\sum_{j}J_{0j}x_{j}^{2}
\end{equation}
where the index $j$ runs over all considered magnetic atoms or 1D unit
cells in the system other than the atom indexed by $0$ and the $x$
component of the displacement vector between the sites $0$ and $j$
is denoted by $x_{j}$. Considering only the edge atoms in the sum
we obtain $D_{\text{edge}}=2308\,\text{meV\AA}^{2}$, while considering
the half ribbon meta magnetic moments we get $D_{\text{half}}=3406\,\text{meV\AA}^{2}$.
Though these values are larger than the one reported in Ref.~\cite{Katsnelson_graphene_ribbon},
they are reasonably similar in magnitude and support the observation that $sp$
magnets might have higher spin stiffness than the conventional $d$
ferromagnets. Based on our calculations we conclude
that the high apparent value of the spin stiffness is caused by the almost
monotonic, unconventional decay of magnetic correlations along the ribbon edge.

\section{Summary}

We presented a computational approach that determines the exchange parameters of
isotropic spin models based on the magnetic force theorem, directly from
\emph{ab initio} calculations using a nonorthogonal basis set to expand the
eigenstates of the system. We demonstrated that the new method accurately
reproduces the Heisenberg interactions of simple metallic bulk ferromagnets
delivered by well--established computational approaches. We studied the
magnetism of two systems based on graphene. For fluorinated graphene we obtained
exchange constants in fairly good agreement with previous calculations using
maximally localized Wannier functions and we confirmed the theoretical
prediction of a 120$^\circ$ N\'eel state. The long-range behavior of the
exchange interactions was found consistent with the electron spectrum of the
system around the Fermi level. For zigzag graphene nanoribbons we found that the
stiffness constant derived from the exchange constants  is consistent with
previous estimates based on total energy differences of twisted spin configurations.
We also found an unconventional $\exp(-\sqrt{r/\delta})$-like decay of the
interaction. Understanding this exotic behavior poses a challenge for further
investigations. The SIESTA code can easily handle large nano-scale systems of
high chemical complexity, therefore we are convinced that the presented method is
a very efficient tool for the analysis and design of novel hybrid nano-structures
hosting exotic magnetic patterns.

\begin{acknowledgments}
This research was supported by the Hungarian Scientific Research Fund
(NKFIH) under Projects No. K115575, No. PD124380, No. K108676, No.
K115608, and No. FK124723. L. O. also acknowledges support from to
the National Quantum Technologies Program NKP-375 2017-00001 of the
NRDI Office of Hungary and from the Bolyai and Bolyai+ Scholarships of the
Hungarian Academy of Sciences. L. S., L. U. and A. D. are grateful
for financial support by the BME-Nanonotechnology FIKP grant of EMMI
(BME FIKP-NAT). J. F. acknowledges funding by MINECO through grant
FIS2015-63918-R.
\end{acknowledgments}

\appendix

\section{Exchange interactions in an orthogonal basis}

In this section we give a detailed derivation of the LKAG formula
for $J_{ij}$ \cite{Lichtenstein_original,Katsnelson_Jij_TBcorr}
to be evaluated by density functional theoretical calculations performed
in an orthogonal tight binding basis. Andersen's force theorem\cite{andersen}
states that if the system is in its ground state then the change of
the total energy $E_{\text{tot}}$ due to a small variation in the
external potential can be directly linked to small variations in the
Kohn--Sham energy $E_{\text{KS}}$ calculated at fixed density without
the need for further self-consistent calculations. The force theorem
thus provides us with a computationally inexpensive way to obtain
response functions.

Neglecting relativistic effects and longitudinal spin fluctuations,
the energy of a spin system is usually mapped to a classical Heisenberg
model,
\begin{equation}
\mathcal{H}=-\frac{1}{2}\sum_{i\ne j}J_{ij}\,\vec{e}_{i}\vec{e}_{j}\,.
\end{equation}
Consider now a ferromagnetic ground state where all spins point in
the same direction $\vec{e}_{0}$ that has the energy
\begin{equation}
E_{0}=-\frac{1}{2}\sum_{i\ne j}J_{ij}.
\end{equation}
If a single spin located at site $i$ is excited to $\vec{e}_{i}\neq\vec{e}_{0}$,
the energy of this single-spin excitation is given by
\begin{equation}
\delta E(\vec{e}_{i})=E(\vec{e}_{i})-E_{0}=\left(1-\vec{e}_{i}\vec{e}_{0}\right)\sum\limits _{k\left(\neq i\right)}J_{ik}\;.
\end{equation}
The energy of a two-site excitation, $\vec{e}_{i}\neq\vec{e}_{0}$
and $\vec{e}_{j}\neq\vec{e}_{0}$ $\left(i\neq j\right)$, can be
expressed as
\begin{align}
 & \delta E(\vec{e}_{i},\vec{e}_{j})=E(\vec{e}_{i},\vec{e}_{j})-E_{0}=\left(1-\vec{e}_{i}\vec{e}_{j}\right)J_{ij}\nonumber \\
 & +\left(1-\vec{e}_{i}\vec{e}_{0}\right)\sum\limits _{k\left(\neq i,j\right)}J_{ik}+\left(1-\vec{e}_{j}\vec{e}_{0}\right)\sum\limits _{k\left(\neq i,j\right)}J_{jk}\,\nonumber \\
 & =\delta E(\vec{e}_{i})+\delta E(\vec{e}_{j})-\left(\vec{e}_{i}-\vec{e}_{0}\right)\left(\vec{e}_{j}-\vec{e}_{0}\right)J_{ij}\,.
\end{align}
The interaction energy between spins $i$ and $j$, $E_{ij}^{\mathrm{int}}$,
is then defined by
\begin{align}
E_{ij}^{\mathrm{int}} & =\delta E(\vec{e}_{i},\vec{e}_{j})-\delta E(\vec{e}_{i})-\delta E(\vec{e}_{j})\label{eq:eint}\\
 & =-J_{ij}\,\delta\vec{e}_{i}\delta\vec{e}_{j}\;,
\end{align}
with $\delta\vec{e}_{i/j}=\vec{e}_{i/j}-\vec{e}_{0}$.

Calculating the energy cost of appropriate local perturbations we
can thus extract the classical $J_{ij}$ parameters from
\emph{ab initio} Green's function methods. Applying Lloyd's formula \cite{Zeller_Lloyd_formula}
in the spirit of the force theorem the energy cost of a perturbation
$\delta\hat{V}$ can be cast in terms of the Green's operator (resolvent)
\begin{equation}
\hat{G}(z)=\left(z\hat{I}-\hat{H}\right)^{-1}\label{eq:green_operator_definition}
\end{equation}
as
\begin{equation}
\delta E_{\text{KS}}=\frac{1}{\pi}\int\limits _{-\infty}^{\varepsilon_\text{F}}\mathrm{d}\varepsilon\,\text{ImTr}\ln\left(\hat{I}-\delta\hat{V}\hat{G}(\varepsilon)\right),
\end{equation}
where $\varepsilon_\text{F}$ is the Fermi energy, $\hat{I}$ is the identity
operator and $\hat{H}$ corresponds to a Hamiltonian which self-consistently
determines $E_{\text{tot}}$. Assuming now that $\delta\hat{V_{i}}$
and $\delta\hat{V_{j}}$ are operators that describe the local perturbations
corresponding to spin rotations at sites $i$ and $j$, respectively,
and using the identity
\begin{equation}
\hat{I}-\hat{A}-\hat{B}=(\hat{I}-\hat{A})(\hat{I}-(\hat{I}-\hat{A})^{-1}\hat{A}\hat{B}(\hat{I}-\hat{B})^{-1})(\hat{I}-\hat{B}),
\end{equation}
we obtain
\begin{align}
 & \ln(\hat{I}-\delta\hat{V}_{i}\hat{G}-\delta\hat{V}_{j}\hat{G})=\nonumber \\
 & \ln(\hat{I}-\delta\hat{V}_{i}\hat{G})+\ln(\hat{I}-\delta\hat{V}_{i}\hat{G})+\ln(\hat{I}-\hat{T}_{i}\hat{G}\hat{T}_{i}\hat{G}),
\end{align}
where the scattering operator $\hat{T}_{i}$ is defined as
\begin{equation}
\hat{T}_{i}=\delta\hat{V}_{i}\left(\hat{I}-\hat{G}\delta\hat{V}_{i}\right)^{-1}\,.
\end{equation}

In the spirit of Eq.~\eqref{eq:eint}, the interaction energy for
the two-site perturbation is then given by
\begin{equation}
E_{ij}^{\mathrm{int}}=\frac{1}{\pi}\int\limits _{-\infty}^{\varepsilon_\text{F}}\mathrm{d}\varepsilon\text{ImTr}\ln\left[\hat{I}-\hat{T}_{i}\hat{G}(\varepsilon)\hat{T}_{j}\hat{G}(\varepsilon)\right].\label{eq:interaction_energy_two_scatterer_with_t}
\end{equation}
Since we are interested in small perturbations around the ground state,
we can safely use the Born approximation, $\hat{T}_{i}\approx\delta\hat{V}_{i}$,
and we also expand the logarithm as $\ln(1-x)\approx-x$, thus Eq.~\eqref{eq:interaction_energy_two_scatterer_with_t}
reduces to
\begin{equation}
E_{ij}^{\mathrm{int}}=-\frac{1}{\pi}\int\limits _{-\infty}^{\varepsilon_\text{F}}\mathrm{d}\varepsilon\,\text{ImTr}\left[\delta\hat{V}_{i}\hat{G}(\varepsilon)\hat{V}_{j}\hat{G}(\varepsilon)\right].\label{eq:interaction_energy_two_scatterer}
\end{equation}
Note that so far we have not considered anything specific about the
perturbation operators $\hat{V}_{i/j}$.

Within the tight binding (TB) scheme a matrix representation of $\hat{H}$
is used in an orthogonal basis of localized atomic-like wavefunctions
centered at sites of the lattice. Thus, the basis functions are labeled
by lattice sites $n$, composite angular momentum indices $L=\left(\ell,m\right)$
and the spin index $s=\pm1/2$ (or $\uparrow$ and $\downarrow$).
As what follows, we will note matrices of the entire site-angular
momentum-spin space with bold-face letters, double and single underlines
will denote block matrices in common angular momentum-spin space and
in only angular momentum space, respectively:
\begin{equation}
\mathbf{H}=\left\{ \underline{\underline{H}}_{nn^{\prime}}\right\} =\left\{ \underline{H}_{ns,n^{\prime}s^{\prime}}\right\} =\left\{ H_{nLs,n'L^{\prime}s^{\prime}}\right\} \;,
\end{equation}
\begin{equation}
H_{nLs,n^{\prime}L^{\prime}s^{\prime}}=H_{nL,n^{\prime}L^{\prime}}^{c}\delta_{ss^{\prime}}+\vec{H}_{nL,n^{\prime}L^{\prime}}\vec{\sigma}_{ss^{\prime}}\;,
\end{equation}
with the Pauli matrices $\vec{\sigma}$. For simplicity, we shall
assume that
\begin{equation}
\vec{H}_{nL,n^{\prime}L^{\prime}}=0\quad\mathrm{for}\quad n\neq n^{\prime}
\end{equation}
and spin dependence applies only to the site-diagonal blocks of the
Hamiltonian,
\begin{equation}
\vec{H}_{nL,nL^{\prime}}=H_{nL,nL^{\prime}}^{s}\,\vec{e}_{n}\;.\label{eq:Hs}
\end{equation}
When the spin is aligned parallel to the $z$ axis, the form of the
local Hamiltonian is
\begin{equation}
\underline{\underline{H}}_{nn}=\left(\begin{array}{cc}
\underline{H}_{nn}^{\uparrow} & 0\\
0 & \underline{H}_{nn}^{\downarrow}
\end{array}\right)\;,
\end{equation}
thus
\begin{align}
\underline{H}_{nn}^{c/s} & =\frac{1}{2}\left(\underline{H}_{nn}^{\uparrow}\pm\underline{H}_{nn}^{\downarrow}\right)\;.
\end{align}
In case of a ferromagnetic (in general, collinear) magnetic configuration
of the host with a magnetic orientation $\vec{e}_{0}$,
\begin{equation}
\underline{\underline{H}}_{nn^{\prime}}=\underline{H}_{nn^{\prime}}^{c}I+\underline{H}_{nn^{\prime}}^{s}\,\vec{e}_{0}\vec{\sigma}\;,
\end{equation}
where $I$ denotes the unit matrix in spin space. The corresponding
matrix representation of the Green's function is of the same form,
\begin{equation}
\underline{\underline{G}}_{nn^{\prime}}=\underline{G}_{nn^{\prime}}^{c}I+\underline{G}_{nn^{\prime}}^{s}\,\vec{e}_{0}\vec{\sigma}\;.\label{eq:GFmatrix}
\end{equation}

According to Eq.~\eqref{eq:Hs} the change of the Hamiltonian due
to local spin rotations is given by 
elements defined as
\begin{equation}
\underline{\underline{\delta\hat{V}_{i}}}_{nn^{\prime}}=\delta_{in}\delta_{in^{\prime}}\underline{H}_{ii}^{s}\,\delta\vec{e}_{i}\vec{\sigma}\;,\label{eq:single_spin_excitation_orthogonal}
\end{equation}
where $\underline{H}_{ii}^{s}$ denotes the angular momentum representation
of the spin-dependent part of the Hamiltonian confined to the site
$i$. In order to calculate the interaction energy of two spins in
Eq.~\eqref{eq:interaction_energy_two_scatterer} we evaluate the
trace by substituting Eqs.~\eqref{eq:GFmatrix} and \eqref{eq:single_spin_excitation_orthogonal}:
\begin{align}
 & \text{Tr}\left[\delta\hat{V}_{i}\hat{G}\delta\hat{V}_{j}\hat{G}\right]=\nonumber \\
 & \mathrm{Tr}_{Ls}\left(\underline{H}_{ii}^{s}\left(\delta\vec{e}_{i}\vec{\sigma}\right)\left[\underline{G}_{ij}^{c}I+\underline{G}_{ij}^{s}\left(\vec{e}_{0}\vec{\sigma}\right)\right]\right.\\
 & \quad\quad\quad\quad\left.\underline{H}_{jj}^{s}\left(\delta\vec{e}_{j}\vec{\sigma}\right)\left[\underline{G}_{ji}^{c}I+\underline{G}_{ji}^{s}\left(\vec{e}_{0}\vec{\sigma}\right)\right]\right)\;,\nonumber
\end{align}
where $\mathrm{Tr}_{Ls}$ denotes the trace of a matrix in both angular
momentum and spin space. Using the algebraic properties of Pauli matrices
the traces can easily be evaluated in spin space yielding
\begin{align}
 & \text{Tr}\left[\delta\hat{V}_{i}\hat{G}\delta\hat{V}_{j}\hat{G}\right]=\nonumber \\
 & 2\mathrm{Tr}_{L}\left[\underline{H}_{ii}^{s}\underline{G}_{ij}^{c}\underline{H}_{jj}^{s}\underline{G}_{ji}^{c}\right.\nonumber \\
 & \quad\quad\left.-\underline{H}_{ii}^{s}\underline{G}_{ij}^{s}\underline{H}_{jj}^{s}\underline{G}_{ji}^{s}\right]\delta\vec{e}_{i}\delta\vec{e}_{j}\nonumber \\
 & +4\mathrm{Tr}_{L}\left[\underline{H}_{ii}^{s}\underline{G}_{ij}^{s}\underline{H}_{jj}^{s}\underline{G}_{ji}^{s}\right]\,\left(\delta\vec{e}_{i}\vec{e}_{0}\right)\left(\delta\vec{e}_{j}\vec{e}_{0}\right)\nonumber \\
 & +2i\mathrm{Tr}_{L}\left[\underline{H}_{ii}^{s}\underline{G}_{ij}^{s}\underline{H}_{jj}^{s}\underline{G}_{ji}^{c}\right.\nonumber \\
 & \quad\quad\quad-\left.\underline{H}_{ii}^{s}\underline{G}_{ij}^{c}\underline{H}_{jj}^{s}\underline{G}_{ji}^{s}\right]\vec{e}_{0}\left(\delta\vec{e}_{i}\times\delta\vec{e}_{j}\right)\,,\label{eq:GVGV}
\end{align}
where $\mathrm{Tr}_{L}$ denotes trace in angular momentum space only.
For infinitesimal rotations 
$\delta\vec{e}_{i}\perp\vec{e}_{0}$, 
therefore, the second term will be neglected. The third term can be
shown to vanish in the present non-relativistic collinear magnetic
case. Due to time-reversal symmetry, the tight-binding basis can be
chosen by unitary transformation such that $\underline{H}_{ij}^{s/c}=\left(\underline{H}_{ji}^{s/c}\right)^{T}$,
consequently also $\underline{G}_{ij}^{s/c}=\left(\underline{G}_{ji}^{s/c}\right)^{T}$,
thus
\begin{align}
 & \mathrm{Tr}_{L}\left[\underline{H}_{ii}^{s}\underline{G}_{ij}^{c}\underline{H}_{jj}^{s}\underline{G}_{ji}^{s}\right]=\nonumber \\
 & =\mathrm{Tr}_{L}\left[\left(\underline{G}_{ji}^{s}\right)^{T}\left(\underline{H}_{jj}^{s}\right)^{T}\left(\underline{G}_{ij}^{c}\right)^{T}\left(\underline{H}_{ii}^{s}\right)^{T}\right]\nonumber \\
 & =\mathrm{Tr}_{L}\left[\underline{G}_{ij}^{s}\underline{H}_{jj}^{s}\underline{G}_{ji}^{c}\underline{H}_{ii}^{s}\right]\nonumber \\
 & =\mathrm{Tr}_{L}\left[\underline{H}_{ii}^{s}\underline{G}_{ij}^{s}\underline{H}_{jj}^{s}\underline{G}_{ji}^{c}\right]\;,
\end{align}
that indeed cancels the third contribution to \eqref{eq:GVGV}.

Thus, the interaction of two spins can indeed be written as
\begin{equation}
E_{ij}^{\mathrm{int}}=-J_{i,}\,\delta\vec{e}_{i}\delta\vec{e}_{j}\;,
\end{equation}
with
\begin{align}
J_{ij} & =\frac{2}{\pi}\int\limits _{-\infty}^{\varepsilon_{\mathrm{F}}}\mathrm{d}E\,\text{Im}\mathrm{Tr}_{L}\left[\underline{H}_{ii}^{s}\underline{G}_{ij}^{c}\underline{H}_{jj}^{s}\underline{G}_{ji}^{c}\right.\;\\
 & \quad\quad\quad\quad\quad\quad\quad\quad\quad\quad-\left.\underline{H}_{ii}^{s}\underline{G}_{ij}^{s}\underline{H}_{jj}^{s}\underline{G}_{ji}^{s}\right].\nonumber
\end{align}

Rewriting $\underline{G}_{ij}^{c/s}$ in terms of $\underline{G}_{ij}^{\uparrow/\downarrow}$
the above expression can be reduced to
\begin{equation}
J_{ij}=\frac{2}{\pi}\int\limits _{-\infty}^{\varepsilon_{\mathrm{F}}}\mathrm{d}\varepsilon\,\text{Im}\mathrm{Tr}_{L}\left[\underline{H}_{ii}^{s}\underline{G}_{ij}^{\uparrow}\underline{H}_{jj}^{s}\underline{G}_{ji}^{\downarrow}\right].\label{eq:Lichtenstein_TB}
\end{equation}
This expression is the celebrated LKAG formula \cite{Lichtenstein_original,Katsnelson_Jij_TBcorr}.
It is important to note that if the magnetic orientations at site $i$ and $j$
are opposite in sign, as happens in case of collinear antiferromagnetic
configurations, the $J_{ij}$ obtained from Eq.~\eqref{eq:Lichtenstein_TB} should
be changed in sign as the spin channels at the two sites are reversed with respect
to each other.

\section{Some identities in a nonorthogonal basis}

Here we review useful identities related to nonorthogonal bases, some
of them discussed in Ref.~\cite{Palacios_nonorthogonal}. Using these
identities we then generalize Eq.~\eqref{eq:Lichtenstein_TB} to
nonorthogonal bases.

A basis formed by states $\left\{ \left|i\right\rangle \right\} $
is not orthogonal if its elements have finite overlap
\begin{equation}
S_{ij}=\left\langle i\right.\left|j\right\rangle .
\end{equation}
In practice real valued basis functions are chosen, therefore the overlap matrix
is symmetric. The inverse of the overlap matrix $\boldsymbol{S}$
defines the dual basis $\left|\tilde{i}\right\rangle $ as
\begin{equation}
\left|\tilde{i}\right\rangle =\sum_{j}(\boldsymbol{S}^{-1})_{ij}\left|j\right\rangle ,
\end{equation}
whose elements are orthogonal to the original basis,
\begin{equation}
\left\langle i\right.\left|\tilde{j}\right\rangle =\delta_{ij}.
\end{equation}
The expansion of a general operator $\hat{A}$ with respect to basis
$\left\{ \left|i\right\rangle \right\} $ is defined as
\begin{equation}
\hat{A}=\sum_{ij}\left|i\right\rangle \tilde{A}_{ij}\left\langle j\right|,
\end{equation}
while the matrix elements in the original basis can be expressed as
\begin{equation}
A_{pq}=\left\langle p\right|\hat{A}\left|q\right\rangle =\sum_{ij}\left\langle p\right.\left|i\right\rangle \tilde{A}_{ij}\left\langle j\right.\left|q\right\rangle .
\end{equation}
Obviously, the expansion coefficients $\tilde{A}_{ij}$ are the matrix
elements of the operator in the dual basis,
\begin{equation}
\tilde{A}_{ij}=\left\langle \tilde{i}\right|\hat{A}\left|\tilde{j}\right\rangle \,.
\end{equation}

As what follows we shall denote the matrix of an operator $\hat{A}$
with respect to the nonorthogonal basis with $\boldsymbol{A}$, while
the matrix in the dual basis will be denoted by $\boldsymbol{\tilde{A}}$.
Note that these two matrices are connected by the overlap matrix
$\boldsymbol{S}$ as
\begin{equation}
\boldsymbol{A}=\boldsymbol{S}\boldsymbol{\tilde{A}}\boldsymbol{S}.
\end{equation}

The trace of an operator is calculated with the help of an orthogonal
basis $\left\{ \left|\alpha\right\rangle \right\}$,
\begin{align}
\text{Tr}\hat{A} & =\sum_{\alpha}\left\langle \alpha\right|\hat{A}\left|\alpha\right\rangle \nonumber \\
 & =\sum_{\alpha ij}\left\langle \alpha\right.\left|i\right\rangle \tilde{A}_{ij}\left\langle j\right.\left|\alpha\right\rangle \nonumber \\
 & =\sum_{\alpha ij}\left\langle j\right.\left|\alpha\right\rangle \left\langle \alpha\right.\left|i\right\rangle \tilde{A}_{ij}\nonumber \\
 & =\sum_{ij}S_{ji}\tilde{A}_{ij}=\text{Tr}(\boldsymbol{S}\boldsymbol{\tilde{A}})\,.
\end{align}
Next we consider matrix elements and traces of operator products.
The trace of a simple product gives
\begin{align}
\text{Tr}(\hat{A}\hat{B}) & =\sum_{\alpha}\left\langle \alpha\right|\hat{A}\hat{B}\left|\alpha\right\rangle \nonumber \\
 & =\sum_{\alpha ijpq}\left\langle \alpha\right.\left|i\right\rangle \tilde{A}_{ij}\left\langle j\right.\left|p\right\rangle \tilde{B}_{pq}\left\langle q\right.\left|\alpha\right\rangle \nonumber \\
 & =\text{Tr}(\boldsymbol{S\tilde{A}S\tilde{B}})\,,
\end{align}
which generalizes to
\begin{equation}
\text{Tr}\prod_{i}\hat{A}_{i}=\text{Tr}\prod_{i}(\boldsymbol{S\tilde{A}_{i}})\,.
\end{equation}
The matrix element of a simple product is expressed as
\begin{align}
\left\langle k\right|\hat{A}\hat{B}\left|l\right\rangle  & =\sum_{ijpq}\left\langle k\right.\left|i\right\rangle \tilde{A}_{ij}\left\langle j\right.\left|p\right\rangle \tilde{B}_{pq}\left\langle q\right.\left|l\right\rangle \\
 & =\left(\boldsymbol{S\tilde{A}S\tilde{B}S}\right)_{kl},\nonumber
\end{align}
that can be generalized to
\begin{equation}
\left\langle k\right|\prod_{i}\hat{A}_{i}\left|l\right\rangle =\left(\boldsymbol{S}\prod_{i}\left(\boldsymbol{\tilde{A}_{i}S}\right)\right)_{kl}.
\end{equation}
Using the Taylor expansion of an operator function,
\begin{equation}
f(\hat{A})=\sum_{n}f_{n}\hat{A}^{n},
\end{equation}
the corresponding trace gives
\begin{align}
\text{Tr}f(\hat{A}) & =\sum_{n}f_{n}\text{Tr}\hat{A}^{n}\nonumber \\
 & =\sum_{n}f_{n}\text{Tr}(\boldsymbol{S\tilde{A}})^{n}\nonumber \\
 & =\text{Tr}f(\boldsymbol{S\tilde{A}}),
\end{align}
while for the respective matrix elements we obtain
\begin{align}
\left\langle k\right|f(\hat{A})\left|l\right\rangle  & =\sum_{n}f_{n}\left\langle k\right|\hat{A}^{n}\left|l\right\rangle \nonumber \\
 & =\sum_{n}f_{n}\left(\boldsymbol{S}(\boldsymbol{\tilde{A}S})^{n}\right)_{kl}\nonumber \\
 & =\left(\boldsymbol{S}f(\boldsymbol{\tilde{A}S})\right)_{kl}\nonumber \\
 & =\left(f(\boldsymbol{S\tilde{A}})\boldsymbol{S}\right)_{kl}.
\end{align}
This identity formally applies to the inverse of an operator,
\begin{align}
\left\langle k\right|\hat{A}^{-1}\left|l\right\rangle  & =\left((\boldsymbol{S\tilde{A}})^{-1}\boldsymbol{S}\right)_{kl}\nonumber \\
 & =\left(\boldsymbol{\tilde{A}}^{-1}\boldsymbol{S}^{-1}\boldsymbol{S}\right)_{kl}\nonumber \\
 & =\left(\boldsymbol{\tilde{A}}^{-1}\right)_{kl},\label{eq:inverse_relation}
\end{align}
but it can be also rigorously proved based on $\hat{A}^{-1}\hat{A}=\hat{I}$.
This means the matrix of the inverse of an operator is the inverse
matrix of the expansion coefficient of the operator.

The density of states $\varrho(E)$ is related to the resolvent $\hat{G}(z)$
in Eq.~\eqref{eq:green_operator_definition} as
\begin{align}
\varrho(\varepsilon) & =-\frac{1}{\pi}\lim_{\delta\rightarrow+0}\text{Im}\text{Tr}\,\hat{G}(\varepsilon+i\delta)\\
 & =-\frac{1}{\pi}\lim_{\delta\rightarrow+0}\text{Im}\text{Tr}(\boldsymbol{S\tilde{G}}(\varepsilon+i\delta))\,.
\end{align}
Using the relation Eq.~\eqref{eq:inverse_relation}, the matrix of
the expansion coefficients of the resolvent reads as
\begin{equation}
\boldsymbol{\tilde{G}}(z)=(z\boldsymbol{S}-\boldsymbol{H})^{-1},
\end{equation}
thus we can express the density of states with the help of the overlap
matrix $\boldsymbol{S}$ and the matrix of matrix elements $\boldsymbol{H}$
as
\begin{equation}
\varrho(\varepsilon)=-\frac{1}{\pi}\lim_{\delta\rightarrow0}\text{Im}\text{Tr}\left[\boldsymbol{S}((\varepsilon+i\delta)\boldsymbol{S}-\boldsymbol{H})^{-1}\right].
\end{equation}
Furthermore, the trace of the product of operators in the interaction
energy for two-site perturbations, Eq.~\eqref{eq:interaction_energy_two_scatterer},
can be calculated as 
to the correlation kernel for two operators $\hat{V_{1}}$ and $\hat{V_{2}}$
defined by
\begin{align}
 & \text{Tr}[\hat{V}_{1}\hat{G}(z)\hat{V}_{2}\hat{G}(z)]=\nonumber \\
 & =\text{Tr}[\boldsymbol{S}\boldsymbol{\tilde{V}}_{1}\boldsymbol{S}\boldsymbol{\tilde{G}}(z)\boldsymbol{S}\boldsymbol{\tilde{V}}_{2}\boldsymbol{S}\boldsymbol{\tilde{G}}(z)]\nonumber \\
 & =\text{Tr}[\boldsymbol{V}_{1}\boldsymbol{\tilde{G}}(z)\boldsymbol{V}_{2}\boldsymbol{\tilde{G}}(z)]\nonumber \\
 & =\text{Tr}[\boldsymbol{V}_{1}(z\boldsymbol{S}-\boldsymbol{H})^{-1}\boldsymbol{V}_{2}(z\boldsymbol{S}-\boldsymbol{H})^{-1}].\label{eq:two_impurity_correlator}
\end{align}

\section{Exchange interactions in a nonorthogonal basis}

In this section we discuss a pragmatic approximation to treat local
spin rotations in a nonorthogonal basis leading to the generalization
of the formula for $J_{ij}$ \eqref{eq:Lichtenstein_TB} derived for
orthogonal basis. Restricting our discussion to collinear magnetic
systems, it is natural to choose a basis where the site and orbital
degrees of freedom form the nonorthogonal part of the basis, while
the basis functions are eigenvectors of the spin operator projected
to the orientation of the magnetization. That is we consider the basis
$\left|\sigma\right\rangle \otimes\left|iL\right\rangle $ with the
property
\begin{align}
\left(\left\langle \sigma\right|\otimes\left\langle iL\right|\right)\left(\left|\sigma'\right\rangle \otimes\left|jL'\right\rangle \right) & =\left\langle \sigma\right.\left|\sigma'\right\rangle \left\langle iL\right.\left|jL'\right\rangle \\
 & =\delta_{\sigma,\sigma'}S_{iL,jL'},\nonumber
\end{align}
where $i$ and $j$ denote lattice sites, $L$ and $L'$ stand for
orbital degrees of freedom and $\sigma$, $\sigma^\prime$ label the eigenvectors
of the spin operator.

Let us define the local perturbation operator as
\begin{equation}
\delta\hat{V}_{i}=\hat{P}_{i}^{\dagger}\left(\hat{O}_{G}^{\dagger}\hat{H}\hat{O}_{G}-\hat{H}\right)\hat{P}_{i}\ ,\label{eq:local_spin_rotation}
\end{equation}
where $\hat{H}$ is a Hamiltonian whose matrix elements have been
calculated by some self-consistent scheme, $\hat{O}_{G}$ describes
a global rotation of the spin degrees of freedom around direction
$\vec{n}$ with angle\textbf{ }$\varphi$
\begin{equation}
\hat{O}_{G}=\text{e}^{-\frac{1}{2}\text{i}\,\vec{n}\vec{\sigma}\,\varphi}\otimes\hat{I}_{L},
\end{equation}
and $\hat{P}_{i}$ is a projector built up from all orbital degrees
of freedom associated with site $i$:
\begin{align}
\hat{P}_{i} & =\hat{I}_{S}\otimes\sum_{L}\left|iL\right\rangle \langle\tilde{iL}\!\mid\,.
\end{align}
The identity operators $\hat{I}_{L}$ and $\hat{I}_{S}$ act on all
orbital degrees of all atomic positions and in spin space, respectively.
Note that this direct Hermitian projection does not project to a subspace
with integer dimension \cite{Palacios_nonorthogonal}. The operator
$\delta\hat{V}_{i}$ has the convenient property that its matrix elements
are only finite between orbitals located at site $i$, and are equal
to the matrix elements of the Hamiltonian rotated globally in spin
space relative to the reference Hamiltonian, $\delta\hat{H}=\hat{O}_{G}^{\dagger}\hat{H}\hat{O}_{G}-\hat{H}$.

Since the global spin rotation and local projection act independently
in the local perturbation, Eq.~\eqref{eq:local_spin_rotation}, the
evaluation of formula \eqref{eq:two_impurity_correlator} follows
the steps as for the orthogonal basis. Thus, the expression of the
Liechtenstein formula is readily generalized to nonorthogonal bases:
\begin{equation}
J_{ij}=\frac{2}{\pi}\int_{-\infty}^{\varepsilon_\text{F}}\text{d}\varepsilon\,\text{ImTr}[\underline{H}_{ii}^{s}\underline{\tilde{G}}_{ij}^{\uparrow}(\varepsilon)\underline{H}_{jj}^{s}\underline{\tilde{G}}_{ji}^{\downarrow}(\varepsilon)]\,,
\end{equation}
with the actual expressions of the above matrices in the nonorthogonal
basis.


 \bibliographystyle{apsrev}
\bibliography{refs_submit}

\end{document}